\documentclass[conference]{IEEEtran}
\IEEEoverridecommandlockouts
\usepackage{cite}
\usepackage{amsmath,amssymb,amsfonts}
\usepackage{algorithmic}
\usepackage{graphicx}
\usepackage{textcomp}
\usepackage{xcolor}
\usepackage{hyperref}

\usepackage{makecell}
\usepackage{tabularx}
\usepackage{threeparttable}
\usepackage{booktabs}
\newcolumntype{Y}{>{\centering\arraybackslash}X}
\usepackage{enumitem}

\usepackage{xcolor}
\definecolor{codegreen}{rgb}{0.0,0.7,0}
\definecolor{codegray}{rgb}{0.5,0.5,0.5}
\definecolor{codepurple}{rgb}{0.58,0,0.82}
\definecolor{backcolour}{rgb}{0.95,0.95,0.92}
\definecolor{yellowish}{rgb}{1.0, 0.98, 0.95}
\definecolor{codeblue}{rgb}{0.25,0.5,0.9}

\usepackage{caption}
\usepackage{listings}
\def\BibTeX{{\rm B\kern-.05em{\sc i\kern-.025em b}\kern-.08em
    T\kern-.1667em\lower.7ex\hbox{E}\kern-.125emX}}
\begin{document}

\title{IFTT-PIN: A Self-Calibrating PIN-Entry Method}

\author{
\IEEEauthorblockN{Kathryn McConkey\IEEEauthorrefmark{1},
Talha Enes Ayranci\IEEEauthorrefmark{1},
Mohamed Khamis\IEEEauthorrefmark{1}, 
Jonathan Grizou\IEEEauthorrefmark{1}}
\IEEEauthorblockA{\IEEEauthorrefmark{1}School of Computing Science, University of Glasgow, Glasgow, United Kingdom}
\IEEEauthorblockA{\IEEEauthorrefmark{1}\{TalhaEnes.Ayranci, Mohamed.Khamis, Jonathan.Grizou\}@glasgow.ac.uk}
}

\maketitle

\begin{abstract}
Personalising an interface to the needs and preferences of a user often incurs additional interaction steps. In this paper, we demonstrate a novel method that enables the personalising of an interface without the need for explicit calibration procedures, via a process we call self-calibration. A second-order effect of self-calibration is that an outside observer cannot easily infer what a user is trying to achieve because they cannot interpret the user's actions. To explore this security angle, we developed IFTT-PIN (If This Then PIN) as the first self-calibrating PIN-entry method. When using IFTT-PIN, users are free to choose any button for any meaning without ever explicitly communicating their choice to the machine. IFTT-PIN infers both the user’s PIN and their preferred button mapping at the same time. This paper presents the concept, implementation, and interactive demonstrations of IFTT-PIN, as well as an evaluation against shoulder surfing attacks. Our study (N=24) shows that by adding self-calibration to an existing PIN entry method, IFTT-PIN statistically significantly decreased PIN attack decoding rate by ca. 8.5 times (p=1.1e-9), while only decreasing the PIN entry encoding rate by ca. 1.4 times (p=0.02), leading to a positive security-usability trade-off. IFTT-PIN's entry rate significantly improved 21 days after first exposure (p=3.6e-6) to the method, suggesting self-calibrating interfaces are memorable despite using an initially undefined user interface. Self-calibration methods might lead to novel opportunities for interaction that are more inclusive and versatile, a potentially interesting challenge for the community. A short introductory video is available at \url{https://youtu.be/pP5sfniNRns}.
\end{abstract}


\section{Introduction} \label{1_introduction}

Personalising a user interface to the needs and preferences of a user often incurs additional interaction steps for the user. For example, a user might be required to navigate menus to define what action to attach to a button, or an expert might be required to configure an interface to a user's needs prior to interaction. Those calibration procedures are important because an interface cannot always be used in the same way by everyone, for example due to sensory or motor impairments.

In this paper, we demonstrate a new method, called self-calibration, that enables the personalising of an interface without the need for explicit calibration procedures. Under self-calibration, a user can interact with a machine using their preferred interaction style right from the start, and the machine can nonetheless infer what the user is trying to achieve. Self-calibration is thus an important property of interactive systems that might contribute in reducing usability barriers.

A second-order effect of self-calibration is that an outside observer cannot easily infer what a user is trying to achieve, simply because they cannot understand the user's actions which are personalised from the start. We chose this security angle as a way to demonstrate self-calibration capabilities via a PIN entry task. This might not be the best application of self-calibration - which we argue later remains to be invented - but it is the best way we have found to explain and allow users to experience the properties of a self-calibrating interfaces.

We developed IFTT-PIN (If This Then PIN) based on the PIN-entry method presented by Roth et al. in  \cite{roth04}, and later iterated upon with methods like SwinPIN \cite{vonzezschwitz15}, which allow users to enter the PIN of their choice via an elimination process. In \cite{roth04}, users click on a button whose color is the same as their digit. IFTT-PIN is adding self-calibration to \cite{roth04}, which means that buttons do not have any color assigned to them at the start of the interaction. Users are free to define the color of each button in their mind and use them as such without informing the interface prior to interaction. After a few iterations, IFTT-PIN can infer both the digit the user had in mind and the color of each button used. IFTT-PIN can thus effectively calibrate itself to each user at interaction time, demonstrating a new interactive experience.

We evaluated IFTT-PIN as a potential defense mechanism against shoulder surfing attacks, where malicious observers try to infer a user password by watching their actions 'over their shoulder’. We present results from a user study (N=24) carried out to determine the usability, security, and motor memorability of IFTT-PIN as an authentication method compared to the traditional PIN-entry method.

We first present our implementation of a Roth et al. interface \cite{roth04}, explain how self-calibration has been implemented to create IFTT-PIN and offer readers to experience IFTT-PIN via a web-based demonstrators. We then review shoulder-surfing literature and present key results from our user study on IFTT-PIN. We conclude by identifying limitations and promising research directions for self-calibrating interfaces.

For the remainder of this paper, we refer to the traditional PIN entry method we find on our smartphones or at an ATM as TRAD. Roth et al. \cite{roth04} interface is referred to as ROTH. Our method is referred to as IFTT-PIN.

\section{Contributions} \label{2_contributions}

\begin{itemize}
\item A new PIN entry method, called IFTT-PIN, leveraging self-calibration to allow users to choose the meaning conveyed by each button on-the-fly without calibration.
\item The first demonstration of a self-calibrating interface using discrete user actions (i.e., using buttons).
\item An online interactive demonstrator that allows to explain the self-calibrating paradigm in a few minutes on a computer or smartphone.
\item A novel approach to overwhelm the observer in shoulder surfing scenarios relying on on-the-fly code-pad configuration.
\item A user study quantifying the usability, security, and memorability of IFTT-PIN, and introducing the SUTO score, a metric for quantifying security-usability trade-offs.
\end{itemize}

\section{Understanding ROTH}

Quoting directly from Roth et al. \cite{roth04}: {\itshape “The principal idea is to present the user the PIN digits as two distinct sets e.g., by randomly coloring half of the keys black and the other half white. The user must enter in which set the digit is by pressing either a separate black or white key. Multiple rounds of this game are played to enter a single digit and it is repeatedly played until all digits are entered. The verifier e.g., the automatic teller machine (ATM), determines the entered PIN digits by intersecting the chosen sets.”}.

In other words, ROTH relies on the following logic: {\itshape “If the user pressed the button B, then they indicated that their digit is of color C, thus their digit is among the set of digits currently colored in C.”}.

We re-implemented Roth et al. interface as follows.

\subsection{Interface Design}

\begin{figure}[h]
  \centering
  \includegraphics[width=\linewidth]{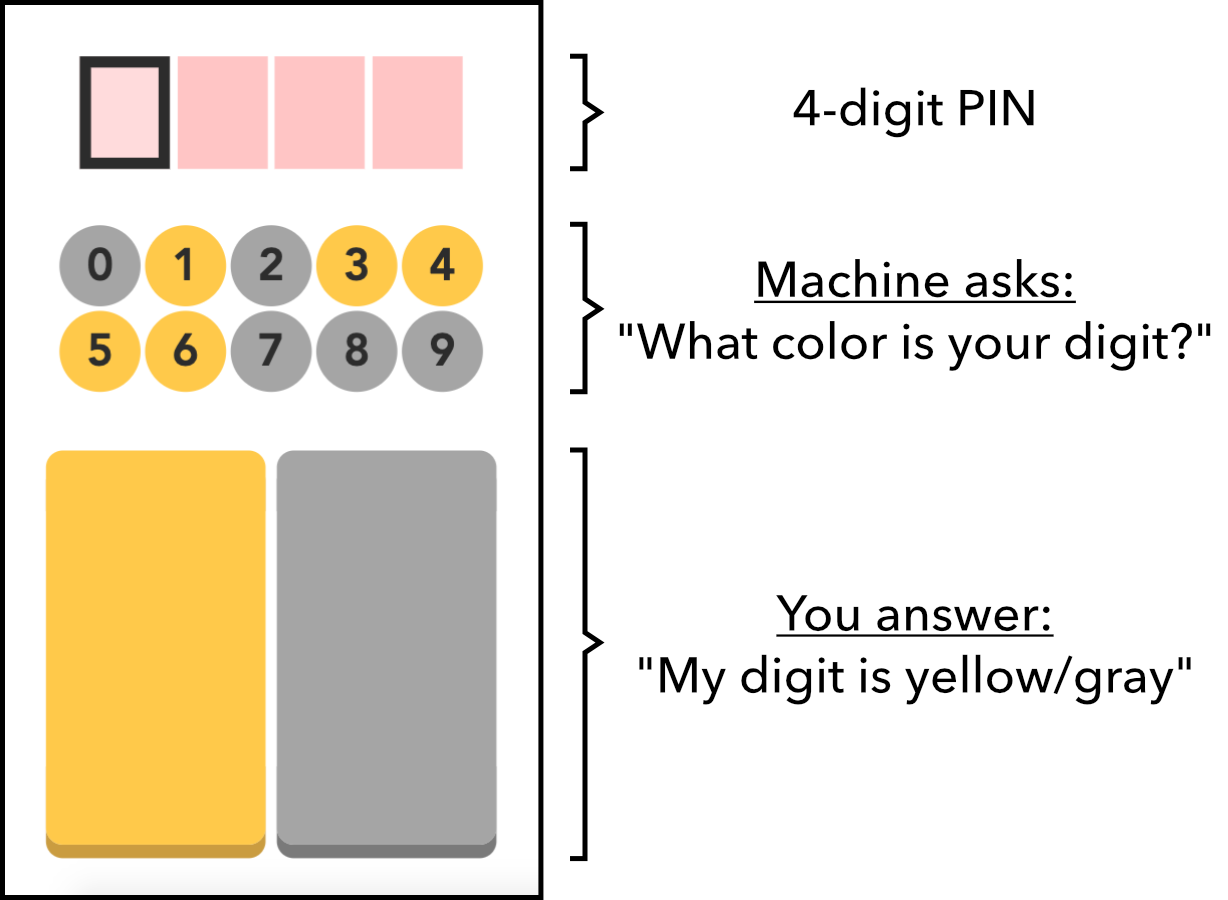}
  \caption{Breakdown of our ROTH interface.}
  \label{fig:breakdown}
\end{figure}

We split our interface into three parts (Figure \ref{fig:breakdown}). The top part displays the PIN, with a black rectangle indicating to the user which of the four digits is currently being entered. The middle part shows all possible digits (from 0 to 9) colored in yellow or gray according to which set they belong to. Think of this section as the machine asking the user: “What color is the digit you want to type?”. The bottom part is dedicated specifically for the user to answer that question. In the case of \cite{roth04}, the user feedback takes place via two colored buttons. Here the left button is yellow, and the right button is gray. Think of this section as the human answering to the machine: “My digit is yellow” or “My digit is gray”.

\subsection{Interaction Vocabulary}

\begin{figure}[h]
  \centering
  \includegraphics[width=\linewidth]{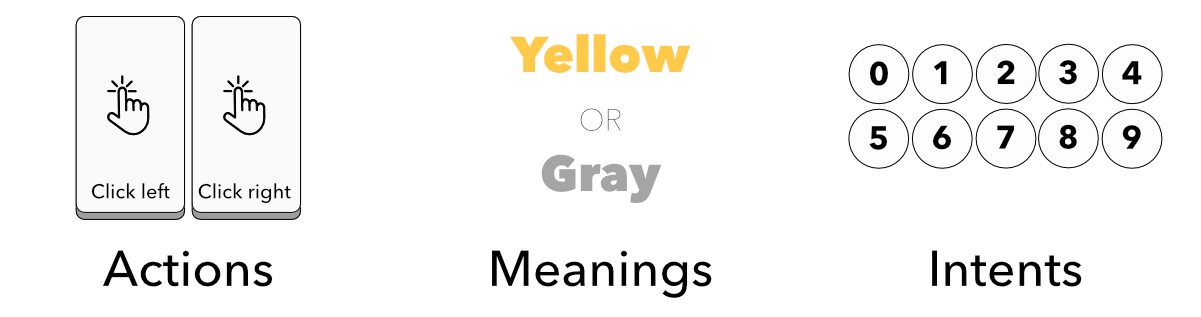}
  \caption{Elements of language. An action conveys a meaning that is used to infer an intent.}
  \label{fig:elements}
\end{figure}

For simplicity, we use a limited interaction vocabulary (Figure \ref{fig:elements}) that helps explain our work in simple steps.

\begin{description}
    \item[Actions] are what the users do in order to convey their meanings. Here the users' actions are to press either the left or the right button
    \item[Meanings] are what the users want to say to the machine, Here the users' meanings are: "My digit is yellow" or "My digit is gray".
    \item[Intents] are what the users want the machine to do, here entering a specific PIN, one digit at a time. 
\end{description}

To put it simply, an \textbf{action} conveys a \textbf{meaning} that is used to infer an \textbf{intent}.

\subsection{Decision Principle} \label{3_4_algorithmic_principle}

\begin{figure}[h]
  \centering
  \includegraphics[width=\linewidth]{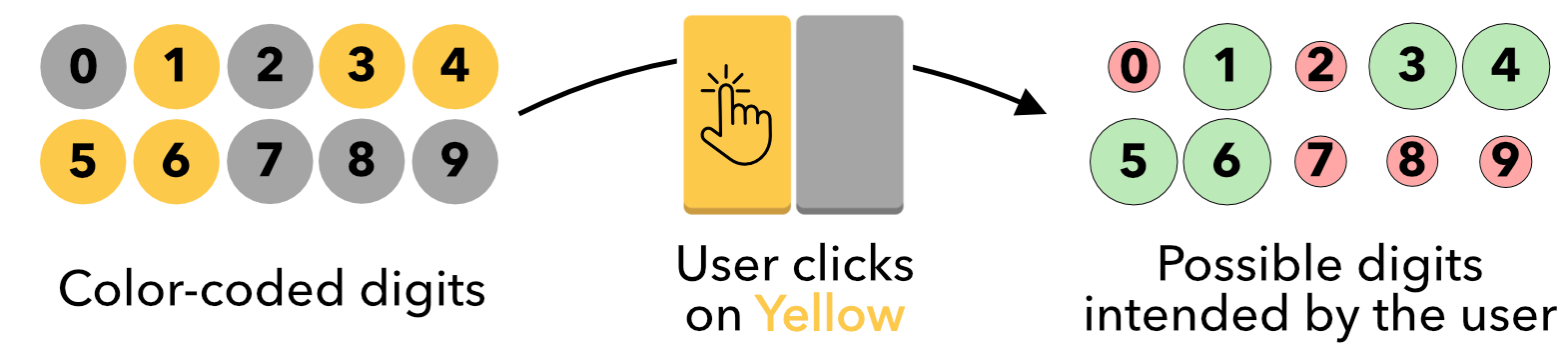}
  \caption{One step of the inference process in ROTH where the action-to-meaning mapping is known.}
  \label{fig:one_step}
\end{figure}

Knowing the color assigned to each button, the interface identifies the digit the user wants to enter by reasoning as follows:{\itshape "If the user pressed the left button (action), then they indicated that their digit is currently yellow (meaning), thus their digit is among the yellow-colored digits and all the gray digits can be discarded (intent)."}, see Figure \ref{fig:one_step}.

By iteratively changing the color applied to each digit, we can narrow the possible digits down to the one the user has in mind. It takes 3 or 4 iterations to identify one digit\footnote{In the best case scenario, we narrow digits down from 10 to 5 to 2 to 1 (3 clicks). Worst case from 10 to 5 to 3 to 2 to 1 (4 clicks).}, hence between 12 and 16 user clicks to enter a 4-digit PIN. Interested readers can refer to Roth et al. pseudocode (Figure 2 in  \cite{roth04}) for a more formal algorithmic description using sets .

\subsection{ROTH Interactive Demonstration} \label{3_5_interactive_demonstration}

We encourage the readers to familiarise themselves with ROTH PIN entry method before moving to the next section. ROTH can be tested online at \url{https://jgrizou.github.io/IFTT-PIN/interaction_1.html} with a walk-through video demonstration at \url{https://youtu.be/6wgOa380uEo}.

\section{Understanding IFTT-PIN}

Having understood and interacted with ROTH, we remind the reader that our contribution with IFTT-PIN lies in the introduction of self-calibration to the ROTH interface.

\subsection{Interface Design}

\begin{figure}[h]
  \centering
  \includegraphics[width=\linewidth]{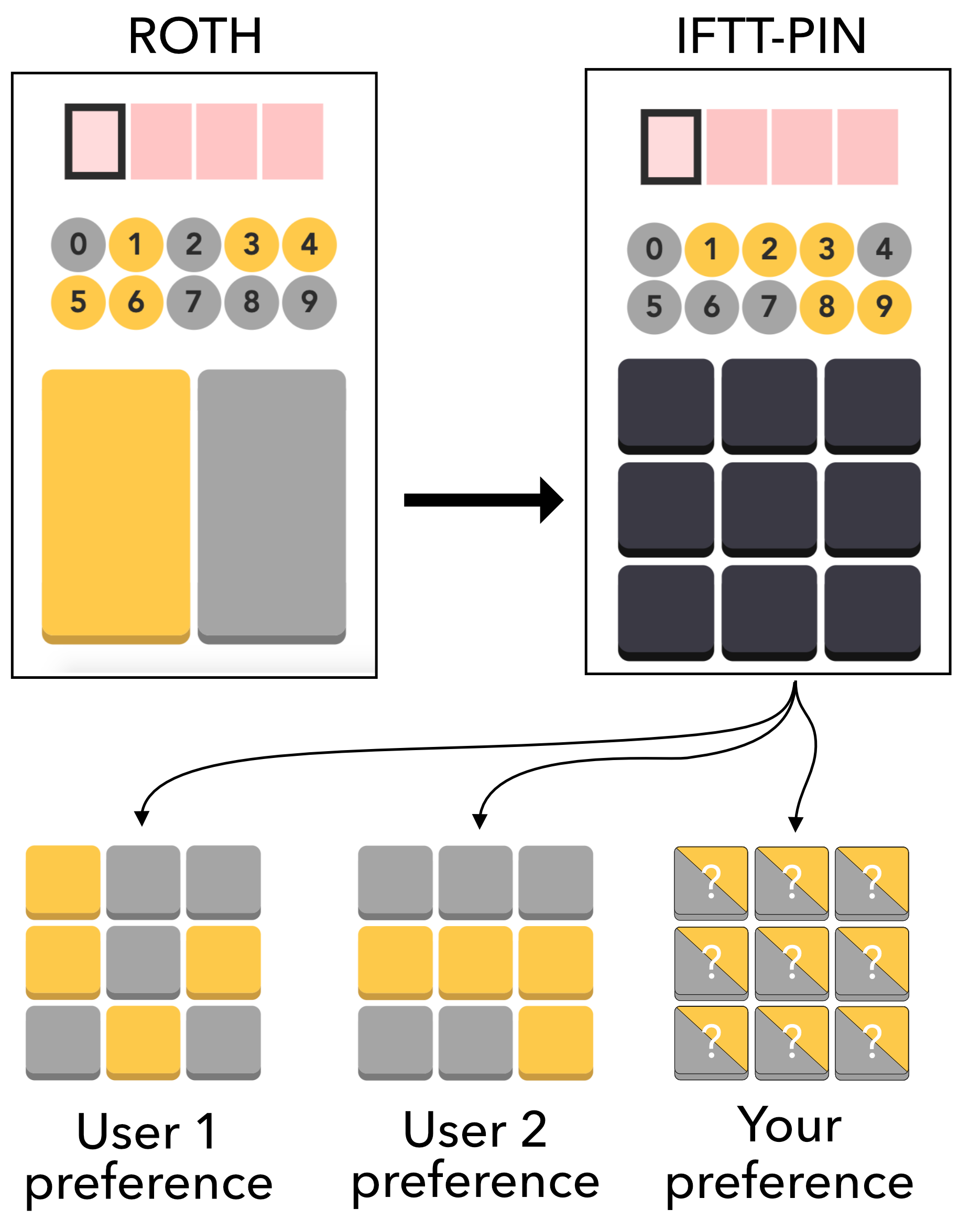}
  \caption{Top: Changes between ROTH and IFTT-PIN. We increased the number of buttons from 2 to 9 to increase the possible color patterns from 2 to 510, and buttons are now undefined (black). Bottom: Examples of choice of button-to-color mapping. At least one button should be assigned to yellow and one to gray.}
  \label{fig:roth_to_iftt}
\end{figure}

To adequately demonstrate the potential of self-calibration, we wanted to allow users to express more varied preferences. We thus increased the number of buttons from 2 to 9, corresponding to an increase in the possible color patterns from 2 to 510\footnote{The number of combinations is $2^N - 2$. The $2^N$ computes the number of possible combinations for 2 colors applied on $N$ buttons. We then add $-2$ because there are $2$ invalid combinations, all yellow or all gray.}. Buttons are now black to indicate that they do not have a color attached to them yet. A user simply decides a color pattern to apply on the buttons -- in their mind and without having to communicate this decision to the device.  The user can start using the buttons that way immediately. For example, the top three buttons could be yellow and all the others gray. See Figure \ref{fig:roth_to_iftt} for more examples.


\subsection{Assumptions}

To explain IFTT-PIN, we first need to understand the assumptions made in ROTH:

\begin{enumerate}
    \item The user's possible intent is one of ten possible digits.\label{assumption:one}
    \item The user's possible meaning is either yellow or gray. \label{assumption:two}
    \item The user can perform one of $N$ actions, pressing one of the $N$ button. \label{assumption:three}
    \item The mapping between the user's actions and their meanings is known. In ROTH, pressing the left button conveys the meaning yellow and the right button conveys the meaning gray. \label{assumption:four}
    \item The user is assumed to be consistent in their usage of the interface. One button can only express one color. \label{assumption:five}
\end{enumerate}

Assumption \ref{assumption:four} is the one we remove in IFTT-PIN, which means that buttons have no predefined colors and the reasoning used in section \ref{3_4_algorithmic_principle} (\textit{“If a user presses the left button (action), then it indicates that their digit is currently yellow (meaning).”}) cannot be used because we do not know the color attached to each button. To work around this problem, we exploit assumption \ref{assumption:five} by explicitly measuring \textit{user's consistency}.

\subsection{Self-calibration Principle}

A user's consistency is something we can observe. If we ask a user to type a specific digit, say 1, and the user presses the same button twice, both when the digit 1 is yellow and again when it is gray, then we can confidently say that the user is being inconsistent with its use of the interface.

But we do not know which digit the user is entering. However, the context given by the PIN entering task and our assumption \ref{assumption:one} ("\textit{The user's possible intent is one of ten possible digits}") constraints the number of possible digits. We can thus formulate ten different hypotheses (one for each digit), and interpret the user’s action according to each hypothetical intent, i.e., each digit a user might be trying to enter. We refer to this process as making \textit{interpretation hypotheses}.

From assumption \ref{assumption:one}, we also know that the user is trying to type only one digit. This implies that the user can only remain consistent with one hypothesis and will, in the long run, invariably breach consistency when their actions are interpreted according to the other hypotheses. By measuring the consistency of the user according to each interpretation hypotheses, we can discard the hypotheses (i.e., the digits) that show a breach of consistency. This combined use of the consistency assumption \ref{assumption:five} and interpretation hypotheses is the basis of self-calibrating systems.

Concretely, IFTT-PIN expands the reasoning used in section \ref{3_4_algorithmic_principle} as follows: {\itshape “If the user is trying to type the digit D, which is currently colored in C, then when they used the button B, they meant the color C”}, and combines it with a test of user's consistency: {\itshape “If, for a particular digit D, the user pressed the same button B to mean two different colors ($C_y$ and $C_g$), then they are not entering the digit D”}. By keeping track of this process for each digit, and over several iterations, IFTT-PIN can identify the digit the user has in mind as the only one for which the interpretation of the user's action remain consistent as per assumption \ref{assumption:five}, that is the only one for which each button pressed was used to mean only one color. This process is easier to understand visually and we have included an illustration in Appendix \ref{appendix} Figure \ref{ifttpinsteps}, as well as a side panel in our demonstrator linked below.

For conciseness, we included algorithmic details, pseudo-code, as well as illustrations of the self-calibration process in Appendix \ref{appendix} for interested readers.

\subsection{IFTT-PIN Interactive Demonstration} \label{subsec:interactive_demonstration}

For the remainder of this paper, we ask readers to familiarise themselves with IFTT-PIN and learn to enter a PIN of their choice using button-to-color mappings of their choice at \url{https://jgrizou.github.io/IFTT-PIN/interaction_2.html}. We developed a version with a side dashboard that displays the history of clicks with respect to each digit, which is available at \url{https://jgrizou.github.io/IFTT-PIN/interaction_2_sidepanel.html}. A walk-through video demonstration can be found at \url{https://youtu.be/t7MQoBnzryQ}. 

\section{Evaluating IFTT-PIN}

We studied user's perception of the usability of our self-calibrating interface, as well as IFTT-PIN effectiveness as a potential protection against shoulder surfing attacks.

\subsection{Shoulder surfing attacks}

Shoulder surfing involves looking over a victim’s shoulder in order to view information on their screen without their explicit consent. Due to sensitivity of content on mobile devices, users tend to be concerned about the privacy of their devices as an individual gaining access may be able to exploit them with malicious intent. 

Smartphones are mainly protected by PINs, which are generally composed of only numbers and are typically short (4-6 digits) for memorability and speed of entry. For mobile devices, predominant PIN lengths are 4 digits which are easily compromised \cite{greene16}. 

Biometric authentication methods, such as fingerprint scanning and facial recognition are the most common alternative to PINs with a study by Markert et al. \cite{markert20} in 2020 showing that 60\% of their participants used biometric methods \cite{markert20}. But Meng et al. has shown that the accuracy of biometric methods is "not stable" \cite{meng15} (e.g., due to sensor failures or poor lighting conditions) and therefore require a knowledge-based backup authentication method, which is often the aforementioned 4 digit numeric PINs\footnote{On Android, users can choose either a PIN or a pattern, and PINs can be over four digits long.}. Thus, biometric authentication "cannot provide universal protection from shoulder surfing attacks" \cite{aviv17}; the extra security they provide is minimal as they are not a replacement for PINs.

The entering of PINs remains the stereotypical shoulder surfing attack scenario and a large body of work focuses on reducing the risk of PIN-based authentication methods that can be securely used in uncontrolled public environments. Common strategies include overwhelming the observer with visual decoys \cite{gugenheimer15,tan05,deluca10}, indirect and/or clustered input  \cite{roth04, deluca10, vonzezschwitz15, vaneekelen13} (e.g., cognitive trap door concept behind \cite{roth04}, combination of gesture and colors with SwiPIN \cite{vonzezschwitz15}), and hidden secondary communication channels \cite{bianchi10, deluca14, khamis16, cueauth} (e.g., tactile feedback with LTF \cite{ku19}, eye-gaze input with GazeTouchCrossPin \cite{ibrahim19} and CueAuthGaze \cite{cueauth}). 

IFTT-PIN offers a novel type of protection mechanism by leveraging self-calibration which removes the need for a pre-established mapping between user's actions and their meanings. The user establishes this mapping on the fly, and can redefine it every time they use the interface, which makes it harder for malicious observers to interpret the user’s actions.

To determine whether self-calibration has the potential to become an effective component of security applications, we designed a user study to determine the usability, security, and memorability of IFTT-PIN compared with the TRAD and ROTH methods, as well as with results from other methods previously reported in the literature.

\subsection{Evaluation Framework}

\begin{table*}[ht]
\caption{How previous research has conducted security and usability evaluations.}
\label{tab:security_usability}
\centering
\begin{tabular}{lcccccc}
    \toprule
    & \multicolumn{2}{c}{Security Evaluation} & \multicolumn{4}{c}{Usability Evaluation} \\
    \cmidrule(lr){2-3} \cmidrule(lr){4-7}
    Method & Human-based & Recording-based & Error Rate & Entry Time & \makecell{NASA TLX} & SUS \\
    \midrule
    SwiPIN \cite{vonzezschwitz15} & x & x & x & x &  & \\
    LTF \cite{ku19} & & & x & x & & \\
    Gaze \cite{ibrahim19} & & x & x & x & & \\
    Graphical \cite{tari06} & x & & x & & & \\
    IOC \cite{roth04} & & x & x & x & & x \\
    CueAuth \cite {cueauth} & x & x & x & x & x & \\
    IFTT-PIN (this paper) & & x & x & x & & x \\
    \bottomrule
\end{tabular}
\end{table*}

All PIN-entry methods we reviewed completed some form of evaluation where both the security and usability were tested, see Table \ref{tab:security_usability}. When evaluating security, studies consist of emulating a shoulder surfing attack and requiring the participant to try to decode PINs. Human-based attacks involve obtaining a user's PIN through direct observation, whereas recording-based attacks involve recording the authentication session to obtain the PIN \cite{binbeshr21}. 

When evaluating usability the majority of studies focused on entry time and error rates, and some included usability questionnaires results, such as the NASA Task Load Index  (NASA TLX) \cite{hart1988development} and the System Usability Score (SUS) \cite{brooke13}. 

In addition, we decided to study IFTT-PIN usability at regular time intervals over 1 month. Our goal was to study motor memorability, not the memorability of the authentication token (here a 4-digit PIN), but rather the ability to remember the workings of an interface after a few days without using it. We also wanted to observe users' performances and potential change of behavior as they become familiar with the interface.

\paragraph{Threat model.} Our attacker is assumed to have a clear, unobstructed, view of the user's device and can record a video of the password being entered.

\paragraph{Variables.} Dependent variables include efficiency, effectiveness, and memorability which will be measured through entry time, error rate and how well participants can remember how to use the interfaces at various time intervals. The independent variables include the device as well as the PIN being entered, with the different conditions being each of the different PIN entry methods (TRAD, ROTH, IFTT-PIN).

\paragraph{Design structure.} We do not expect participants to experience any fatigue, therefore, we can choose a within subject design which means each participant is exposed to all of the conditions.

\paragraph{Time constraints.} To ensure a bounded experimental time, we have constrained security and usability tests to 30 seconds for TRAD, 3 minutes for ROTH, and 5 minutes for IFTT-PIN.

\paragraph{Ordering.} Within a session, we evaluated methods in order of algorithmic complexity: TRAD first, then ROTH, then IFTT-PIN.

\paragraph{Training.} We assumed that all participants have experience with TRAD. For ROTH and IFTT-PIN, all participants watched a short explanatory video explaining how to use the interfaces.

\paragraph{Choice of PIN.} We required participants to select one of three PINs between 1234, 4321 and 2468, which were deemed of similar difficulty. Their chosen PIN had to be used across all authentication methods and could not be a PIN that the participants currently use.

\paragraph{Timeline} On day 0, participants undertook first the security tests, then the usability tests. Participants came back on day 7 and day 21 and repeated the usability tests only.

\paragraph{Comparison to related PIN entry methods} We report our experimental results alongside results extracted from related PIN entry methods listed in Table \ref{tab:security_usability}.

\subsection{Security} We chose recording-based attacks where participants watched a video showing someone entering PINs with a completely unobstructed view, with full control over the videos and were allowed to pause and rewind at will. We measure how many PINs users can correctly decode within a set time frame, as well as the digit decoding rate per minutes.

Only 2 participants were able to decode 1 or 2 digits from IFTT-PIN during the 5 minutes initially planned for. To avoid a flooring effect, we conducted an additional 1-hour long study to determine more accurately the security level of IFTT-PIN with a subset of 7 participants.

\subsection{Usability} We measured the "extent to which a product can be used by specified users to achieve specified goals with effectiveness, efficiency and satisfaction in a specified context of use" \cite{iso} as follows:

\begin{description}
    \item[Effectiveness] is the ``accuracy and completeness with which users achieve specified goals" \cite{tari06}. We measure effectiveness through the error rates when entering PINs.
    \item[Efficiency] is the ``resources expended in relation to the accuracy and completeness with which users achieve goals"\cite{tari06}. We measure this through the time taken to enter PINs as well as the rate of digit entering per minute.
    \item[Satisfaction] is the users being ``free from discomfort with positive attitudes towards the use of the product"\cite{tari06}. Users filled the the System Usability Scale (SUS) questionnaire \cite{brooke13} for each user authentication method. This scale gives us a ``global view of subjective usability" and is known to be a reliable qualitative measure of usability. As a reference, a SUS score of 68 would be considered average \cite{brooke13}.
\end{description}

\subsection{Memorability} We measured ``how well users re-establish proficiency with the system functionality after a time period not using the system"\cite{nielsen12}. We repeated our usability tests with the same participants at three intervals, day 0, day 7 and day 21.

\section{Results} \label{Results}

We invited 24 participants (14 males, 10 females) aged between 18 and 24, of which 7 took part in the extended 1h-long security study on IFTT-PIN. Before completing the study, all participants were presented with information sheets and consent forms and were made aware they could withdraw from the study at any time without any consequences. All participants agreed to participate and filled out demographic questionnaires which also provided questions about knowledge of authentication methods as well as shoulder surfing. 

Using a Likert scale, we determined the IT proficiency level of participants. 71.6\% claimed to be either proficient or highly proficient. 83.3\% of participants felt very comfortable using a mobile phone while the remaining participant felt comfortable. 67\% of participants knew what the term shoulder surfing meant. After being provided with a definition, 42\% of participants said that they had experienced a shoulder surfing attack. 54.2\% of participants felt concerned about shoulder surfing and would be willing to have more secure authentication methods implemented for daily use. One participant had heard about IFTT-PIN previously but did not enter or decode a PIN via IFTT-PIN prior to this study.

\subsection{Security Results} \label{sec:security}

\begin{table}
\begin{threeparttable}
    \caption{\textbf{Security results.} CueAuthGaze has the lowest reported attack success rate (0.05\%) while IFTT-PIN has the lowest per digit decoding rate (0.12 \textit{digits/min}) suggesting an attacker would take longer to decode a PIN entered via IFTT-PIN than via other methods. By adding self-calibration to the ROTH method, IFTT-PIN significantly increased PIN attack decoding rate by ca. 8.5 times (p=1.1e-9)}
    \label{tab:security_results}
    \begin{tabularx}{\columnwidth}{X Y Y}
        \toprule
        Method & \makecell{Attack success \\ (\textit{\%})} & \makecell{Decoding rate \\ (\textit{digits/min)}} \\
        \midrule
        SwiPIN \cite{vonzezschwitz15} & 12.5 & x \\
        LTF \cite{ku19} & x & x \\
        Gaze \cite{ibrahim19} & 14.2 & x  \\
        Graphical \cite{tari06} & 11 & x  \\
        IOC \cite{roth04} & 8.75  & x \\
        CATouch \cite{cueauth} & 74 & 2.31  \\
        CAMidAir \cite{cueauth} & 64 & 2.63  \\
        CAGaze \cite{cueauth} & \textbf{0.05} & 1.47\tnote{c}  \\
        TRAD & 99 & 71.33 (SD=2.26) \tnote{l} \\ 
        ROTH & 8 & 1.03 (SD=0.59) \\ 
        IFTT-PIN (ours) & 3.125 & 
        \textbf{0.12 (SD=0.12)}\tnote{e} \\
        \bottomrule
    \end{tabularx}
    \begin{tablenotes}
        \item[f] {Likely flooring effect due to the very low percentage of participants successfully decoding the PIN, decoding rate is thus likely lower for CueAuthGaze.}
        \item[l] {Lower bound estimate due to ceiling effect during our TRAD experiments.} 
        \item[e] {Based on results from the 1-hour long security experiment.} 
    \end{tablenotes}
\end{threeparttable}
\end{table}

\begin{table*}
\begin{threeparttable}
    \caption{\textbf{Usability results.} IFTT-PIN is the slowest entry method reported with an average entry time of ca. 38 seconds and an average encoding rate of ca. 8 digits per minutes. CueAuthGaze, which scored well on security metrics, has the highest reported PIN entry error rate (17.28\%, ca. 4.5x higher than IFTT-PIN). By adding self-calibration to the ROTH method, IFTT-PIN significantly decreased PIN entry encoding rate by ca. 1.4 times (p=0.02).}
    \label{tab:usability_results}
    \begin{tabularx}{\textwidth}{X Y Y Y} 
        \toprule
        Method & \makecell{PIN Entry Time \\ (\textit{seconds})} & \makecell{PIN Entry Error Rate \\ (\textit{\%)}} & \makecell{Encoding Rate \\ (\textit{digits/min})} \\
        \midrule
        SwiPIN \cite{vonzezschwitz15} & 3.66 (SD = 0.9) & 3.1 & 65.57 \\
        LTF \cite{ku19} & 13.84 & 2 & 17.33 \\
        Gaze \cite{ibrahim19} & 9.5 & x & 25.26 \\
        Graphical \cite{tari06} & x & x & x  \\
        IOC \cite{roth04} & 23.23 & 9 & 10.33 \\
        CueAuthTouch \cite{cueauth} & 3.73 (SD = 0.98) & 6.62 (SD = 5.56\tnote{e}) & 64.34 \\
        CueAuthMidAir \cite{cueauth} & 5.51 (SD = 3.87) & 15.81 (SD = 8.16\tnote{e}) & 43.55 \\
        CueAuthGaze \cite{cueauth} & 26.35 (SD = 22.09) & \textbf{17.28 (SD = 8.45}\tnote{e}\textbf{)} & 9.11 \\
        TRAD & 1.37 (SD = 0.41) & 4.04 (SD = 5.89) & 196.67 (SD = 46.72) \\ 
        ROTH & 24.19 (SD = 3.76) & 1.88 (SD = 2.42) & 10.92 (SD = 1.13) \\ 
        IFTT-PIN (this paper) & \textbf{38.01 (SD = 14.86)} & 3.76 (SD = 3.88) & \textbf{7.91 (SD = 2.24)} \\
        \bottomrule
    \end{tabularx}
    \begin{tablenotes}
        \item[e] SD approximated based on standard deviation information on success rates in \cite {cueauth}. 
    \end{tablenotes}
\end{threeparttable}
\end{table*}

Security results are reported in Table \ref{tab:security_results}. CueAuthGaze has the lowest reported attack success rate (0.05\%) while IFTT-PIN has the lowest per digit decoding rate (0.12 \textit{digits/min}) suggesting an attacker would take longer to decode a PIN entered via IFTT-PIN than via other methods. Further, longer, experiments would be required to quantify the decoding rates of all methods.

A repeated measures ANOVA showed a significant main effect for the entry method on attack decoding rate (F=6088, p=2.5e-78). Post hoc Bonferroni corrected t-tests revealed significant differences between TRAD and ROTH (t=76, p=1.6e-49), TRAD and IFTT-PIN (t=143, p=3.8e-62), and ROTH and IFTT-PIN (t=7.9, p=1.1e-9). Overall, our results confirmed that by adding self-calibration to the ROTH method, IFTT-PIN significantly decreased PIN attack decoding rate by ca. 8.5 times (p=1.1e-9).

We emphasize that the average decoding rate for TRAD is likely higher than the 71.33 \textit{digits/min} reported due to a likely ceiling effect caused by a limited supply of TRAD recorded attacks. All but one participants could decode all the PINs presented to them during the TRAD experiments, hence they might have been able to decode even more PINs within the allocated time.

From our observations and interviews, the dominant decoding strategy for IFTT-PIN was to list out the different numbers in each color-set for each button click. When participants observed a button was clicked multiple times, they started seeking which digit stayed consistently within one color-set and determined that this digit was the correct digit. From there, participants could reverse engineer the color pattern and reconstruct the rest of the PIN.

\subsection{Usability Results} \label{sec:usability}

Usability results are reported in Table \ref{tab:usability_results}. IFTT-PIN is the slowest entry method reported with an average entry time of ca. 38 seconds (M=38.01, SD=14.86) and an average encoding rate of ca. 8 digits per minutes (M=7.91, SD=2.24). CueAuthGaze, which scored well on security metrics, has the highest reported PIN entry error rate (17.28\%), ca. 4.5x higher than IFTT-PIN. Interestinglt, IFTT-PIN has one of the lowest PIN entry error rates of the methods reported in Table \ref{tab:usability_results}. 

A repeated measures ANOVA showed a significant main effect for the entry method on PIN entry time (F=105, p=1.2e-21) with post hoc Bonferroni corrected t-tests revealing significant differences between TRAD and ROTH (t=-29.5, p=4.4e-31), TRAD and IFTT-PIN (t=12, p=2.2e-15), and ROTH and IFTT-PIN (t=-4.4, p=1.8e-4). Encoding rate showed similar level of significance for the entry method effect on encoding rate (F=137, p=9.3e-25) with post hoc Bonferroni corrected t-tests revealing significant differences between TRAD and ROTH (t=13.6, p=2.9e-17), TRAD and IFTT-PIN (t=11.1, p=3.9e-14), and ROTH and IFTT-PIN (t=-2.85, p=0.02). However no significant differences were found between entry method on the PIN entry error rate. Overall, our results confirmed that by adding self-calibration to the ROTH method, IFTT-PIN significantly decreased PIN entry encoding rate by only ca. 1.4 times (p=0.02), which is to be compared with the ca. 8.5 times (p=1.1e-9) decrease in PIN attack decoding rate reported in section \ref{sec:security}.

Metrics reported for IOC have been extracted from the original paper from Roth et al. \cite{roth04} and should match with our user testing of ROTH. We found a close match between IOC and ROTH for PIN entry time (23.23 and 24.19 respectively), but found a surprising discrepancy on the PIN entry error rate (respectively 9\% and 1.88\%) which we have not been able to explain.

We found a surprisingly high error rate for TRAD (M=4.04, SD=5.89) which we attribute to our observation that participants felt competitive and tried to maximise entry speed during TRAD entry tests, despite being instructed against it.

\subsection{Memorability results} \label{lm} \label{trends}

\begin{table}
    \caption{\textbf{Spaced IFTT-PIN results.} The encoding rate for IFTT-PIN has improved significantly between day 0 and day 21 (p=3.6e-6) suggesting that self-calibration is a memorable interaction paradigm. Users could remember how to enter a PIN without needing to remember specific color patterns.}
    \label{tab:memorability}
    \begin{tabular}{cccc}
        \toprule
        IFTT-PIN & Day 0 & Day 7 & Day 21 \\
        \midrule
        \makecell{Encoding Rate \\ (\textit{digits/min})} & \makecell{6.26 \\ (SD=2.00)} &  \makecell{7.47 \\ (SD=2.22)} & \makecell{\textbf{7.91} \\ \textbf{(SD=2.24)}} \\
        \makecell{Error Rate \\ (\textit{\%)}}  & \makecell{5.97 \\ (SD=6.10)} & \makecell{4.41 \\ (SD=5.18)}  & \makecell{\textbf{3.76} \\ \textbf{(SD=3.88)}} \\
        SUS Score & 51.56 & 55.2 & \makecell{\textbf{58.13}} \\ 
        \bottomrule
    \end{tabular} 
\end{table}

Memorability results are reported in Table \ref{tab:memorability}. A repeated measures ANOVA showed a significant main effect for the day of evaluation on encoding rate (F=3.75, p=0.03) with post hoc Bonferroni corrected t-tests revealing significant differences between day 0 and day 7 (t=-4.15, p=1.2e-3), day 0 and day 21 (t=-6.5, p=3.6e-6), but not a significant difference between day 7 and day 21 (t=-1.65, p=0.33). No significant differences were found between evaluation days on the PIN entry error rate. Overall, our results confirmed that a self-calibrating interface like IFTT-PIN is memorable even if not used for a couple of weeks, with users effectively remembering how to enter a PIN without needing to remember specific color patterns.

On day 21, the overall usability was highest for TRAD (96.31), followed by ROTH (79.88), and then IFTT-PIN (58.13). While IFTT-PIN SUS score increased after each testing sessions, it remained relatively low indicating IFTT-PIN remains less usable than established methods and suffer from usability issues \cite{bangor09}.

At a behavioral level, we noticed that many participants started by using all 9 buttons on IFTT-PIN, but on day 21 most participants reduced their usage to 2 or 3 buttons which is consistent with a desire for speed during PIN entry.
 
\subsection{SUTO - Security-Usability Trade-Off}

Security applications face a security-usability trade-off (SUTO). Simply put, a system that is secure and not usable will not be used while a system that is usable and not secure is at risk of attack \cite{cranor04}. 

To quantify the usability vs. security trade-off, we define a SUTO score as the ratio of the rate of digits being entered to the rate of digits being decoded. 

\begin{equation}
    \text{SUTO score} = \frac{\text{Rate of entering}}{\text{Rate of decoding}}
\end{equation}

The SUTO score quantifies how difficult it is for a PIN to be simultaneously observed and decoded. A SUTO score of 1 indicates that the time required to enter the PIN equals the time needed to decode it. A SUTO score above 1 indicates that the attacker would take longer to decode the PIN than to enter it. The higher the SUTO scores the better, within limits of usability.

SUTO scores are reported in Table \ref{tab:suto}. ROTH scored 10.62 and IFTT-PIN 65.91, confirming that adding self-calibration provides an additional protection against shoulder surfing attacks (ca. 8.5 times, see section \ref{sec:security}) without a corresponding decrease in entry time (ca. 1.4 times, see section \ref{sec:usability}), hence firmly improving the security-usability trade-off. To put this into perspective, on average, it would take about 41 minutes to decode a PIN that took 38 seconds to enter using IFTT-PIN -- assuming full access to a video recording.

CueAuthGaze \cite{cueauth}, which scored high on security metrics, has a SUTO score of 6.20 suggesting a less favorable ratio of entering time over decoding time. We will refrain from making conclusion based on this results as the decoding rate we reconstructed for CueAuthGaze might be affected by flooring effect as described in the security section.

\begin{table}
\begin{threeparttable}
    \caption{\textbf{SUTO Scores.} IFTT-PIN has the highest SUTO scores of all methods. Adding self-calibration to ROTH improved the SUTO score by more than 6-fold, suggesting that self-calibration offers interesting properties where security-usability trade-off are important.}
    \label{tab:suto}    
    \begin{tabularx}{\columnwidth}{X Y} 
        \toprule
        Method & \makecell{SUTO score} \\
        \midrule
        SwiPIN \cite{vonzezschwitz15} & x \\
        LTF \cite{ku19} & x \\
        Gaze \cite{ibrahim19} & x \\
        Graphical \cite{tari06} & x \\
        IOC \cite{roth04} & x \\
        CueAuthTouch \cite{cueauth} & 27.85\tnote{r} \\
        CueAuthMidAir \cite{cueauth} & 16.53\tnote{r} \\
        CueAuthGaze \cite{cueauth} & 6.20\tnote{r,i} \\
        TRAD & 2.75\tnote{h} \\ 
        ROTH & 10.62 \\ 
        IFTT-PIN (this paper) & \textbf{65.91} \\
        \bottomrule
    \end{tabularx}
    \begin{tablenotes}
        \item[r] Reconstructed based on information from \cite {cueauth}. 
        \item[i] SUTO score likely underestimated (i.e., lower bound) due to low attack success rate (flooring effect) inducing a likely lower bound estimation of the decoding rate reconstructed in Table \ref{tab:security_results}, longer experiments are required to estimate accurate SUTO score.
        \item[h] Higher bound estimate due to a lower bound estimates of the decoding rate in Table \ref{tab:security_results}, likely value closer to 1 as real-time decoding is feasible.
    \end{tablenotes}
\end{threeparttable}
\end{table}

\section{Limitations and Future Work} \label{8_limitations}

While IFTT-PIN scores the highest on the SUTO score, we must keep in mind that its usability as a daily PIN entry mechanism is limited due to its high entry time (M=38.01) and below average system usability scores (M=58.13). Similarly to other methods, IFTT-PIN might be better positioned for less frequent, but more secure, PIN entry requirements, such as protecting specific sensitive applications on a mobile devices (e.g., a banking app).

In our opinion, interesting follow-up studies would look into the diversity of users’ choice of action-to-meaning mapping for self-calibrating interfaces. The question is whether allowing users to choose how to use an interface effectively leads to more varied and personalized interaction schemes, and whether users change their preferences over time. Studying these preferences for other interaction modalities, such as gestures or voice commands, would add interesting challenges both algorithmic and in terms of interface design.

To improve usability, options to consider includes adding feedback mechanisms (e.g., a way to show how far along users are in the digit entry process), increasing the color-set size to reduce entry time, reducing the number of buttons (9 buttons was confusing to users), and adding a method to account for user mistakes (e.g., an undo last-action button). 

To improve the SUTO score, and as already suggested in Roth et al. \cite{roth04}, we could introduce a probabilistic approach where the user does not need to enter the exact PIN to open the vault, but simply to reduce the set of possible PINs to a small number that includes the correct PIN. An attacker would only be able to recover a set of possible PINs, not knowing which specific PIN is the correct one. Another approach would be to introduce a hidden communication channel that only the user can access. For example, using special glasses that reveal the color assigned to each digit to the user only. Outside observers or cameras would not see the color on the digit, making it impossible to reverse-engineer the PIN.

For interested parties, more studies ``in the wild" would inform how users interact differently with IFTT-PIN in an uncontrolled environment. Repeating previous studies to compute more reliable SUTO scores for other PIN entry methods would also benefit the community.

\section{Conclusion} \label{9_conclusion}

We presented IFTT-PIN, a PIN-entry method conceived as a vehicle to introduce the self-calibration paradigm. IFTT-PIN allows users to enter the PIN of their choice via an elimination process by indicating the color assigned to their digit (yellow or gray). To express their choice, users click on a button whose color is the same as their digit. But buttons do not have any color assigned to them at the start of the interaction. Users are free to define the color of each button in their mind and use them as such without informing the interface. After a few iterations, IFTT-PIN can infer both the digit the user had in mind and the color of each button used. IFTT-PIN can thus effectively calibrate itself to each user at interaction time, demonstrating a new interactive experience.

For this reason, we proposed IFTT-PIN as a novel approach to protect against shoulder surfing attacks. Because the color of each button is not pre-defined and only pre-exists in the user’s mind, it is significantly harder for an attacker to make sense of the user's actions in real time. By adding self-calibration to ROTH, IFTT-PIN statistically significantly decreased PIN attack decoding rate by ca. 8.5 times (p=1.1e-9), while only decreasing the PIN entry encoding rate by ca. 1.4 times (p=0.02), leading to a positive security-usability trade-off with a SUTO score increasing from 10.62 for ROTH to 65.91 for IFTT-PIN. IFTT-PIN's entry rate significantly improved 21 days after first exposure (p=3.6e-6), suggesting self-calibrating interfaces are memorable despite using an initially undefined user interface. 

While it remains unclear how self-calibrating interfaces could be genuinely beneficial in our everyday world, we would like to encourage researchers to investigate what other interfaces and applications could benefit from the proposed self-calibration paradigm - especially towards more inclusive interactive devices where users can express their intention in an intuitive, fluid and personalized way without resorting to explicit calibration procedures. We hope this work will help inspire the community to invent them.

\section*{Acknowledgment}

We extend our gratitude to all the participants of the study, whose involvement was essential to the success of this research. Jonathan Grizou ideated and built IFTT-PIN based on previous research on self-calibrating interactive systems. Kathryn McConkey carried out the user study under the supervision of Jonathan Grizou and Mohamed Khamis. Talha Enes Ayranci helped with data analysis and served as editor for the paper. Mohamed also played a key role in reviewing and further developing the manuscript.

\bibliographystyle{IEEEtran}
\bibliography{bibliography}

\begin{thebibliography}{10}
\providecommand{\url}[1]{#1}
\csname url@samestyle\endcsname
\providecommand{\newblock}{\relax}
\providecommand{\bibinfo}[2]{#2}
\providecommand{\BIBentrySTDinterwordspacing}{\spaceskip=0pt\relax}
\providecommand{\BIBentryALTinterwordstretchfactor}{4}
\providecommand{\BIBentryALTinterwordspacing}{\spaceskip=\fontdimen2\font plus
\BIBentryALTinterwordstretchfactor\fontdimen3\font minus \fontdimen4\font\relax}
\providecommand{\BIBforeignlanguage}[2]{{%
\expandafter\ifx\csname l@#1\endcsname\relax
\typeout{** WARNING: IEEEtran.bst: No hyphenation pattern has been}%
\typeout{** loaded for the language `#1'. Using the pattern for}%
\typeout{** the default language instead.}%
\else
\language=\csname l@#1\endcsname
\fi
#2}}
\providecommand{\BIBdecl}{\relax}
\BIBdecl

\bibitem{roth04}
\BIBentryALTinterwordspacing
V.~Roth, K.~Richter, and R.~Freidinger, ``A pin-entry method resilient against shoulder surfing,'' in \emph{Proceedings of the 11th ACM Conference on Computer and Communications Security}, ser. CCS '04.\hskip 1em plus 0.5em minus 0.4em\relax New York, NY, USA: Association for Computing Machinery, 2004, p. 236–245. [Online]. Available: \url{https://doi.org/10.1145/1030083.1030116}
\BIBentrySTDinterwordspacing

\bibitem{vonzezschwitz15}
\BIBentryALTinterwordspacing
E.~von Zezschwitz, A.~De~Luca, B.~Brunkow, and H.~Hussmann, ``Swipin: Fast and secure pin-entry on smartphones,'' in \emph{Proceedings of the 33rd Annual ACM Conference on Human Factors in Computing Systems}, ser. CHI '15.\hskip 1em plus 0.5em minus 0.4em\relax New York, NY, USA: Association for Computing Machinery, 2015, p. 1403–1406. [Online]. Available: \url{https://doi.org/10.1145/2702123.2702212}
\BIBentrySTDinterwordspacing

\bibitem{greene16}
\BIBentryALTinterwordspacing
K.~K. Greene, J.~Kelsey, and J.~M. Franklin, \emph{Measuring the Usability and Security of Permuted Passwords on Mobile Platforms}, Apr. 2016. [Online]. Available: \url{http://dx.doi.org/10.6028/NIST.IR.8040}
\BIBentrySTDinterwordspacing

\bibitem{markert20}
P.~Markert, D.~V. Bailey, M.~Golla, M.~D{\"u}rmuth, and A.~J. Aviv, ``This pin can be easily guessed: Analyzing the security of smartphone unlock pins,'' in \emph{2020 IEEE Symposium on Security and Privacy (SP)}.\hskip 1em plus 0.5em minus 0.4em\relax IEEE, 2020, pp. 286--303.

\bibitem{meng15}
W.~Meng, D.~S. Wong, S.~Furnell, and J.~Zhou, ``Surveying the development of biometric user authentication on mobile phones,'' \emph{IEEE Communications Surveys \& Tutorials}, vol.~17, no.~3, pp. 1268--1293, 2015.

\bibitem{aviv17}
\BIBentryALTinterwordspacing
A.~J. Aviv, J.~T. Davin, F.~Wolf, and R.~Kuber, ``Towards baselines for shoulder surfing on mobile authentication,'' in \emph{Proceedings of the 33rd Annual Computer Security Applications Conference}, ser. ACSAC '17.\hskip 1em plus 0.5em minus 0.4em\relax New York, NY, USA: Association for Computing Machinery, 2017, p. 486–498. [Online]. Available: \url{https://doi.org/10.1145/3134600.3134609}
\BIBentrySTDinterwordspacing

\bibitem{gugenheimer15}
\BIBentryALTinterwordspacing
J.~Gugenheimer, A.~De~Luca, H.~Hess, S.~Karg, D.~Wolf, and E.~Rukzio, ``Colorsnakes: Using colored decoys to secure authentication in sensitive contexts,'' in \emph{Proceedings of the 17th International Conference on Human-Computer Interaction with Mobile Devices and Services}, ser. MobileHCI '15.\hskip 1em plus 0.5em minus 0.4em\relax New York, NY, USA: Association for Computing Machinery, 2015, p. 274–283. [Online]. Available: \url{https://doi.org/10.1145/2785830.2785834}
\BIBentrySTDinterwordspacing

\bibitem{tan05}
D.~S. Tan, P.~Keyani, and M.~Czerwinski, ``Spy-resistant keyboard: more secure password entry on public touch screen displays,'' in \emph{Proceedings of the 17th Australia Conference on Computer-Human Interaction: Citizens Online: Considerations for Today and the Future}, ser. OZCHI '05.\hskip 1em plus 0.5em minus 0.4em\relax Narrabundah, AUS: Computer-Human Interaction Special Interest Group (CHISIG) of Australia, 2005, p. 1–10.

\bibitem{deluca10}
\BIBentryALTinterwordspacing
A.~De~Luca, K.~Hertzschuch, and H.~Hussmann, ``Colorpin: securing pin entry through indirect input,'' in \emph{Proceedings of the SIGCHI Conference on Human Factors in Computing Systems}, ser. CHI '10.\hskip 1em plus 0.5em minus 0.4em\relax New York, NY, USA: Association for Computing Machinery, 2010, p. 1103–1106. [Online]. Available: \url{https://doi.org/10.1145/1753326.1753490}
\BIBentrySTDinterwordspacing

\bibitem{vaneekelen13}
\BIBentryALTinterwordspacing
W.~A. van Eekelen, J.~van~den Elst, and V.-J. Khan, ``Picassopass: a password scheme using a dynamically layered combination of graphical elements,'' in \emph{CHI '13 Extended Abstracts on Human Factors in Computing Systems}, ser. CHI EA '13.\hskip 1em plus 0.5em minus 0.4em\relax New York, NY, USA: Association for Computing Machinery, 2013, p. 1857–1862. [Online]. Available: \url{https://doi.org/10.1145/2468356.2468689}
\BIBentrySTDinterwordspacing

\bibitem{bianchi10}
\BIBentryALTinterwordspacing
A.~Bianchi, I.~Oakley, V.~Kostakos, and D.~S. Kwon, ``The phone lock: audio and haptic shoulder-surfing resistant pin entry methods for mobile devices,'' in \emph{Proceedings of the Fifth International Conference on Tangible, Embedded, and Embodied Interaction}, ser. TEI '11.\hskip 1em plus 0.5em minus 0.4em\relax New York, NY, USA: Association for Computing Machinery, 2010, p. 197–200. [Online]. Available: \url{https://doi.org/10.1145/1935701.1935740}
\BIBentrySTDinterwordspacing

\bibitem{deluca14}
\BIBentryALTinterwordspacing
A.~De~Luca, M.~Harbach, E.~von Zezschwitz, M.-E. Maurer, B.~E. Slawik, H.~Hussmann, and M.~Smith, ``Now you see me, now you don't: protecting smartphone authentication from shoulder surfers,'' in \emph{Proceedings of the SIGCHI Conference on Human Factors in Computing Systems}, ser. CHI '14.\hskip 1em plus 0.5em minus 0.4em\relax New York, NY, USA: Association for Computing Machinery, 2014, p. 2937–2946. [Online]. Available: \url{https://doi.org/10.1145/2556288.2557097}
\BIBentrySTDinterwordspacing

\bibitem{khamis16}
\BIBentryALTinterwordspacing
M.~Khamis, F.~Alt, M.~Hassib, E.~von Zezschwitz, R.~Hasholzner, and A.~Bulling, ``Gazetouchpass: Multimodal authentication using gaze and touch on mobile devices,'' in \emph{Proceedings of the 2016 CHI Conference Extended Abstracts on Human Factors in Computing Systems}, ser. CHI EA '16.\hskip 1em plus 0.5em minus 0.4em\relax New York, NY, USA: Association for Computing Machinery, 2016, p. 2156–2164. [Online]. Available: \url{https://doi.org/10.1145/2851581.2892314}
\BIBentrySTDinterwordspacing

\bibitem{cueauth}
\BIBentryALTinterwordspacing
M.~Khamis, L.~Trotter, V.~M\"{a}kel\"{a}, E.~v. Zezschwitz, J.~Le, A.~Bulling, and F.~Alt, ``Cueauth: Comparing touch, mid-air gestures, and gaze for cue-based authentication on situated displays,'' \emph{Proc. ACM Interact. Mob. Wearable Ubiquitous Technol.}, vol.~2, no.~4, dec 2018. [Online]. Available: \url{https://doi.org/10.1145/3287052}
\BIBentrySTDinterwordspacing

\bibitem{ku19}
W.-C. Ku and H.-J. Xu, ``Efficient shoulder surfing resistant pin authentication scheme based on localized tactile feedback,'' in \emph{2019 6th IEEE International Conference on Cyber Security and Cloud Computing (CSCloud)/ 2019 5th IEEE International Conference on Edge Computing and Scalable Cloud (EdgeCom)}, 2019, pp. 151--156.

\bibitem{ibrahim19}
\BIBentryALTinterwordspacing
D.~M. Ibrahim and S.~Ambreen, ``Gaze touch cross pin: Secure multimodal authentication using gaze and touch pin,'' p. 777–781, Oct. 2019. [Online]. Available: \url{http://dx.doi.org/10.35940/ijeat.A1381.109119}
\BIBentrySTDinterwordspacing

\bibitem{tari06}
\BIBentryALTinterwordspacing
F.~Tari, A.~A. Ozok, and S.~H. Holden, ``A comparison of perceived and real shoulder-surfing risks between alphanumeric and graphical passwords,'' in \emph{Proceedings of the Second Symposium on Usable Privacy and Security}, ser. SOUPS '06.\hskip 1em plus 0.5em minus 0.4em\relax New York, NY, USA: Association for Computing Machinery, 2006, p. 56–66. [Online]. Available: \url{https://doi.org/10.1145/1143120.1143128}
\BIBentrySTDinterwordspacing

\bibitem{binbeshr21}
\BIBentryALTinterwordspacing
F.~Binbeshr, M.~{Mat Kiah}, L.~Y. Por, and A.~Zaidan, ``A systematic review of pin-entry methods resistant to shoulder-surfing attacks,'' \emph{Computers \& Security}, vol. 101, p. 102116, 2021. [Online]. Available: \url{https://www.sciencedirect.com/science/article/pii/S0167404820303898}
\BIBentrySTDinterwordspacing

\bibitem{hart1988development}
S.~G. Hart and L.~E. Staveland, ``Development of nasa-tlx (task load index): Results of empirical and theoretical research,'' in \emph{Advances in psychology}.\hskip 1em plus 0.5em minus 0.4em\relax Elsevier, 1988, vol.~52, pp. 139--183.

\bibitem{brooke13}
J.~Brooke, ``Sus: a retrospective,'' \emph{J. Usability Studies}, vol.~8, no.~2, p. 29–40, feb 2013.

\bibitem{iso}
\BIBentryALTinterwordspacing
ISO. (1998) Ergonomic requirements for office work with visual display terminals (vdts) — part 11: Guidance on usability. [Online]. Available: \url{https://www.iso.org/obp/ui/#iso:std:iso:9241:-11:ed-1:v1:en}
\BIBentrySTDinterwordspacing

\bibitem{nielsen12}
\BIBentryALTinterwordspacing
J.~Nielsen. (2012, Jan) Usability 101: Introduction to usability. [Online]. Available: \url{https://www.nngroup.com/articles/usability-101-introduction-to-usability/}
\BIBentrySTDinterwordspacing

\bibitem{bangor09}
A.~Bangor, P.~Kortum, and J.~Miller, ``Determining what individual sus scores mean: adding an adjective rating scale,'' \emph{J. Usability Studies}, vol.~4, no.~3, p. 114–123, may 2009.

\bibitem{cranor04}
L.~Cranor and S.~Garfinkel, ``Guest editors' introduction: Secure or usable?'' \emph{IEEE Security \& Privacy}, vol.~2, no.~5, pp. 16--18, 2004.

\bibitem{grizouinteractive14}
J.~Grizou, I.~n. Iturrate, L.~Montesano, P.-Y. Oudeyer, and M.~Lopes, ``Interactive learning from unlabeled instructions,'' in \emph{Proceedings of the Thirtieth Conference on Uncertainty in Artificial Intelligence}, ser. UAI'14.\hskip 1em plus 0.5em minus 0.4em\relax Arlington, Virginia, USA: AUAI Press, 2014, p. 290–299.

\end{thebibliography}

\appendix

\section{Appendix} \label{appendix}

\subsection{Algorithmic Illustration}

\begin{figure*}[h]
  \centering
  \includegraphics[width=1.0\textwidth]{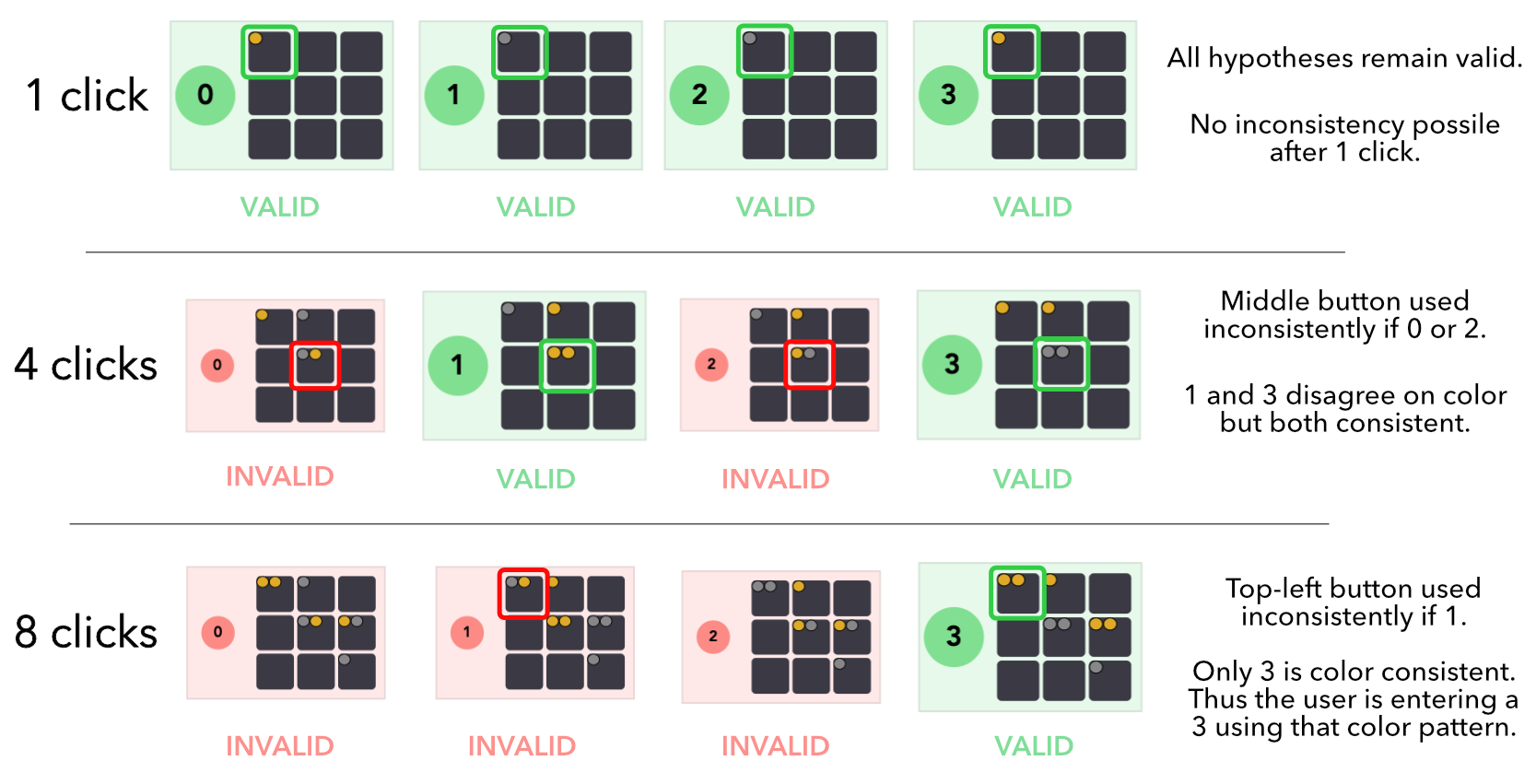}
  \caption{Illustration of inconsistency detection for digits 0 to 3 after one, four, and eight clicks from a typical interaction. After each iteration, a dot is placed on the button pressed by the user and is colored of the same color as was the color of the digit when the button was pressed. Green squares highlight buttons of interest for which hypothesis is consistent. Red ones highlight inconsistencies, meaning the same button would have been used to mean two different colors. Notice how none of the hypotheses share the same button-to-color mapping, yet several mappings can remain consistent for many steps. For example, after 4 clicks, hypothesis 1 and 3 disagree on the color to assign to the middle button. Yet, in both cases, the usage of the button is consistent and thus both hypotheses remain valid.}
  \label{ifttpinsteps}
\end{figure*}

The interpretation hypotheses process is illustrated in Figure \ref{ifttpinsteps} after one, four, and eight clicks from a typical interaction. For visual clarity, we only show the process for digit 0, 1, 2, and 3, as if the user could only enter one of those four digits. 

After one click, the top-left button is marked with a dot because the user just clicked on that button. This dot is yellow for digit 0 and 3, and gray for digit 1 and 2, because, at the time the user clicked on the button, the digits 0 and 3 were yellow, and digits 1 and 2 were gray. After one click, there is no way to know which digit the user is entering, nor what color is assigned to that button, all hypotheses are consistent. 

After four clicks, the middle button has received two clicks from the user. If the user was entering digit 0 and 2, they would have used the middle button to mean alternatively yellow and gray, which would be a breach of the consistency assumption. Thus, the user is not entering digit 0 or 2. For digit 1 and 3, both times the user clicked on the middle button to mean the same color. The middle button would be yellow if the user was typing a 1, and gray if the user was typing a 3, but both options remain consistent at this stage, and the user might well be entering any of them, we just can’t tell yet. 

After eight clicks, the top-left button would have been used for both yellow and gray if the user was entering digit 1. Only our interpretation of the button color according to digit 3 remains fully consistent, thus we can conclude that the user is trying to enter the digit 3. This process is best understood interactively and an interactive demonstration is available on the link shared in section \ref{subsec:interactive_demonstration}).

Once we have identified a digit, we automatically know which color the user is attributing to each button. In other words, the process of entering the first digit under the self-calibration paradigm is calibrating the system for subsequent use by the same user. Within the IFTT-PIN interface, we show this by assigning colors to each button once the interface has identified a new digit. When the user reuses the same buttons to enter a second digit, it will therefore take less iteration than when entering the first digit by falling back to the reasoning from section \ref{3_4_algorithmic_principle}. For buttons that remain undefined (black), we keep using the self-calibration algorithm.

The sequence of colors applied to digits is important to identify to the correct hypothesis as fast as possible. We detail this active learning problem in appendix \ref{subsec:activelearning}.

\subsection{Algorithmic Details}

Measuring consistency requires us to keep track of past interactions and to interpret them in light of each possible digit $(d \in D)$ the user might be intending to enter. Specifically, we keep the history of users button presses $(b \in B)$ and digit colors $(c \in C)$ and run the following reasoning for each digit: {\itshape “If the user is trying to type the digit $d$, then when they used button $b$, they meant color $c$.”} For each digit $d$, the algorithm is building a history of $(b,c)$ pairs representing samples from the action-to-meaning mapping if the user was entering $d$. For a given digit $d$, this historical sequence can be written as $H_N^d = \{ (b_i, c_i \mid d) \mid i = 1, \ldots, N \}$ where  $b_i$ and $c_i  | d$  represent respectively the button pressed by the user and the color of the digit $d$ at iteration $i$ out of $N$ iteration performed so far.

Under our consistency assumption, a user cannot use the same button to express more than one color. For each digit and button, this reasoning unfolds as: {\itshape “If a user is trying to enter digit $d$, and used the button $b$ to mean both $c^{yellow}$ and $c^{gray}$, then the user would have been inconsistent and is thus not trying to enter $d$.”}  This is the core reasoning behind IFTT-PIN.

Assuming no human error, Boolean logic is sufficient to test for consistency. Specifically, if the number of colors assigned to a button is greater that one, then the user is being inconsistent. We can define $H_N^{d,b}={ ( c_i  | d)}$ where $b=b_i$ and $i= 1,\ldots,N$ as a subset of $H_N^d$  containing only the history of colors from $H_N^d$ that are linked with the button $b$. If, for a given digit $d$, for all $b$, the cardinality (number of unique elements in a set, here number of unique colors) of $H_N^{d,b}$ is greater than 1, then the user is inconsistent. Using mathematical notation, for a given $d$, inconsistency is defined as $\exists b \in B,|H_N^{d,b}| > 1$, and, conversely, consistency can be written as  $\forall b \in B,|H_N^{d,b} | < 1$. In turn, the consistency assumption could be expressed using the uniqueness quantification notation as $\exists! d \in D,(\forall b \in B,|H_N^{d,b} | \leq 1)$, which can be read as: ``there is exactly one digit for which, for all buttons, the number of colors assigned to each button was at most one''. 

This uniqueness is satisfied only once the user has “interacted enough” with the interface. It is valid at the limit when $N\rightarrow \infty$ and enough “variety of color pattern” has been displayed on the digit. In other words, only one unique hypothesis will eventually remain consistent. In practice, in most cases, the user digit can be identified after only a few clicks as the reader will be able to test via the interactive demo linked in section \ref{subsec:interactive_demonstration}. 

In summary, to identify the user digit, IFTT-PIN measures consistency for all digits after each iteration. Once only one digit satisfies the consistency requirement, we know it is the one the user has in mind. A pseudocode of this process is summarized in Listing \ref{lst:pseudocode} using a Pythonic syntax.

\begin{figure*}[b]
\noindent\hspace*{1cm}
\begin{minipage}{0.9\textwidth}
\begin{lstlisting}[caption={Pseudocode for IFTT-PIN. The history of button presses and digit colors are stored in a variable called \textit{historyPerDigit} (L12-13). A test of consistency is performed (L14-25) and all consistent digits are added to the \textit{consistentDigitis} list (L27-28). The process stops when only one consistent digit remains in the list (L8), which is the digit the user has in mind (see L30-31). The implementation details for \textit{applyColorToDigits()} (L9) and \textit{waitForUserInput()} (L10) may vary per application, see section \ref{subsec:activelearning}).}, label={lst:pseudocode}]
nDigit = 10; nButton = 9
# init history of interaction H_N^{d,b}
historyPerDigit = [ []*nButton ]*nDigit # empty at start
# init list of possible digits the user has in mind
consistentDigits = list(range(nDigit)) # all digits are possible at start

# repeat until only one user digit is consistent
while (len(consistentDigits) > 1):
    digitColors = applyColorToDigits() # return list of color for each digit
    buttonPressed = waitForUserInput() # return button pressed by the user
    # store history of color per button per digit
    for digit in range(nDigit):
        historyPerDigit[digit][buttonPressed].append(digitColors[digit])
    # test for consistency for each digit
    consistentDigits = [] # empty list of consistent digit
    for digit in range(nDigit):
        # tmp variable, assumed consistent unless prooven otherwise
        isDigitConsistent = True
        # check consistency button per button
        for button in range(nButton):
            # compute the set of unique color attached to this button
            colorSet = set(historyPerDigit[digit][button])
            # inconsistent if more than one color in the set
            if (len(colorSet) > 1):
                isDigitConsistent = False
        # if digit consistent, we add it to the consistent list
        if (isDigitConsistent):
            consistentDigits.append(digit)

# consistentDigits[0] is the digit the user has in mind
# historyPerDigit[consistentDigits[0]] contains button-to-color mapping info



\end{lstlisting}
\end{minipage}
\end{figure*}


\subsection{Active Choice of Colors} \label{subsec:activelearning}

In  \cite{roth04}, the sequence of color was pre-defined and ensured that the digit could be uniquely identified. In IFTT-PIN, we are actively deciding which color to assign to each digit at each iteration according to the history of interaction and in order to identify the user intent as fast as possible. For self-calibration problems, an active planning strategy has been proposed in \cite{grizouinteractive14} which we implemented in IFTT-PIN. We added an additional constraint that each colored set of digits must be balanced at each iteration, in other words, there must always be five yellow digits and five gray digits. The implementation details are beyond the scope of this paper, but our code will be made available online for inspection (not linked here for blind review).

\end{document}